Friends for Free:
Self-Organizing Artificial Social Networks for Trust and Cooperation[1]

David Hales (dave@davidhales.com), Stefano Arteconi (arteconi@cs.unibo.it)
Dept. of Computer Science,
University of Bologna, Italy
August 2005

**Abstract**

By harvesting friendship networks from e-mail contacts or instant message "buddy lists" Peer-to-Peer (P2P) applications can improve performance in low trust environments such as the Internet. However, natural social networks are not always suitable, reliable or available. We propose an algorithm (SLACER) that allows peer nodes to create and manage their own friendship networks.

We evaluate performance using a canonical test application, requiring cooperation between peers for socially optimal outcomes. The Artificial Social Networks (ASN) produced are connected, cooperative and robust - possessing many of the disable properties of human friendship networks such as trust between friends (directly linked peers) and short paths linking everyone via a chain of friends.

In addition to new application possibilities, SLACER could supply ASN to P2P applications that currently depend on human social networks thus transforming them into fully autonomous, self-managing systems.

**Keywords:** Trust, Cooperation, P2P, Social Networks, Self-Organization.

**1. Introduction**

It is well known that human social relationships form a social network that spans the entire globe. These networks form a small-world topology, in which highly connected clusters of mutual friends are linked to other clusters by individuals that form "social bridges" or hubs. The upshot of this, as has been dramatically demonstrated (Travers et al 1969), is that it is often possible to find a short chain of friends of friends to anyone on the globe – you are probably linked to the President of the United States of America by no more than about six hops in your friendship network.

Friendship networks have several desirable properties, they tend to be cooperative, we trust our friends, and they support a number of social functions that we need to achieve in everyday life – we can ask friends for favors or cooperatively solve a difficult problem with work colleagues. We can turn to friends for good advice or discuss with them confidential issues. Furthermore, they are constructed and maintained in a completely distributed way – there's no central authority decreeing who should be part of your friendship network.

---

[1] This work partially supported by the EU within the 6th Framework Program under contract 001907 (DELIS).

It is for these reasons that, increasingly, these networks are being "harvested" and used to solve tricky collaborative tasks in computer networks. For example, to improve P2P message routing (Martini et al 2005) - because friends computers can be trusted to pass on messages - or to support collaborative spam filtering (Kong et al 2004) – if my friends computer says "it's spam" then my computer believes its spam.

## 1.1 The Need for Artificial Social Networks

However, these systems import *existing* social networks that are built and maintained in the everyday social world by people. This takes time and they will tend to be biased towards the social goals of the people involved. For example, in the office the majority of your e-mail contacts are likely to be work colleagues but on a home computer the bias might be towards friends and family. In either case, who is to say that the trust established between friends in a human social context *necessarily* translates into useful trust in some computer application? My friends computer may say "it's spam" but his machine might be incorrectly configured or infected with malicious code that makes its advice untrustworthy. Personally, I do not make friends based on their eagerness to download security patches!

Our approach is to import *network formation processes* so applications can maintain their *own* social networks tuned to their *own* goals. Put another way, we get computers to make their own trusted friends - "friends for free".

Researchers have been trying to solve these problems for some time, namely, how to get cooperation and trust between computers sharing open networks when there is no central authority or control. One effect of the current lack of such trust are nasty problems that plague the internet that most of us probably know all too well; proliferation of e-mail spam, the spread of malicious viruses, worms, trojans, malware and spyware.

## 1.2 Overview

We introduce a simple algorithm (SLACER) which, when executed in the nodes of a P2P network, self-organizes the network into a robust trusted ASN with small-world characteristics and high cooperation. The algorithm is an extension of the SLAC[2] algorithm (Hales 2004, Hales et al 2005) based on the "tagging" approach originating from Computational Sociology (Riolo et al 2001, Hales 2000) that has been shown to support high-levels of cooperation without the need for central control, reciprocity or other evaluation mechanisms. It is based on some simple rules of social behavior observed in human societies.

---

[2] Which bares comparison to SLIC (Sun et al 2004).

We tested SLACER's ability to produce *cooperative and connected* ASN by having nodes play the Prisoner's Dilemma (PD) game. In the PD there is always a temptation for players to betray their partner for personal gain, so this is a good test of trust or "healthy friendship" between nodes.

SLACER produced high trust ASN that connected the entire population of nodes with almost all nodes reachable from any other via chains of cooperative nodes. The ASN were robust and scalable even though nodes act in a selfish individualistic way – If they see another node outperforming them, then they try to copy their behavior and link to their friends – rather like a greedy social climber.

We briefly introduce P2P overlay networks, describe the SLACER algorithm in detail, and then present the PD test application used for evaluation. After this we give simulation results and indicate a number of practical application areas for which, we believe, the SLACER algorithm provides a sound basis.

**2. Peer-to-Peer Overlay Networks**

The target infrastructures we have in mind for the SLACER algorithm are unstructured peer-to-peer (P2P) overlay networks. In a P2P overlay network there is a population of nodes, typically processes situated within a physical network, which maintain links to other nodes (often called their neighbors). P2P applications, like Skype[3] or BitTorrent[4] implement these to provide services. When we say a node is "linked" in an overlay net we do not mean there is a physical link between them, but rather an address or label is stored in the nodes that allows them to communicate. So for example, an overlay net running on the Internet might represent links as IP addresses. Each node can store some number of IP's of other nodes in the network. An overlay net therefore allows us to abstract away from the physical infrastructure supporting the communication between nodes. We simply assume that such an infrastructure layer exists. Hence, if two nodes are linked in the overlay this indicates nothing about their physical network proximity, rather, the overlay describes a logical topology that runs "on-top" of any physical network that can provide the necessary communication services such as reliable rooting between all nodes.

In the work presented here, we also make a further abstraction; we assume that such an overlay network can be modeled as an undirected graph. In such a graph, all links (termed edges) between nodes (termed vertices) are undirected. In the context of a P2P we interpret this as a constraint indicating that all links between nodes are symmetric – if a node *i* links to a node *j* then node *j* also links to node *i*. This means we only represent bidirectional communication links in our overlay and we assume the underlying physical network can provide this service.

---

[3] www.skype.com
[4] www.bittorrent.com

One valuable property of the overlay net abstraction is that rewiring nodes or changing the topology of the network is a logical process in which nodes simply drop, copy or exchange symbolic links. It is therefore not costly to maintain highly dynamic network topologies at the overlay layer. This makes the manipulation of dynamic networks in real time feasible.

**3. The SLACER Algorithm**

The SLACER algorithm is an extension of the SLAC algorithm (Hales 2004) – a simple node rewiring algorithm that produces cooperation when nodes behave selfishly. It's a Selfish Link-based Adaptation for Cooperation algorithm (SLAC). SLACER follows the same general approach but is more conservative because it maintains some old links rather than rewiring all links – it's a Selfish Link-based Adaptation for Cooperation Excluding Rewiring algorithm (SLACER). These acronyms actually do mean things!

We assume that peer nodes can change their strategy (i.e. change the way they behave at the application level) and drop and make links to nodes they know about. It is also assumed nodes have the ability to discover other nodes randomly from the entire population, compare their performance against those nodes in some way and copy their links and strategies[5]. SLACER implements a simple local adaptation rule: Nodes try to use their abilities to selfishly increase their own performance (or utility) in a greedy and adaptive way by changing their links and strategy. They do this by copying nodes that appear to be performing better and by making randomized changes with low probability.

Figure 1 shows the pseudocode. Over time, nodes engage in some application task and generate some measure of utility $U$. This utility is a numeric value that each node needs to calculate based on the specifics of the particular application domain. For example, this might be number of files downloaded, jobs processed or an inverse measure of spyware infections over some period. The higher the value of $U$ the better the node believes it is performing in its target domain.

---

[5] For a discussion of how we might deal with malicious noise in the exchange of information between nodes see the conclusion.

| Active thread: | Passive thread: |
|---|---|
| i ← this node<br>do forever:<br>Engage in application task<br>update i.Utility<br>Periodically (compare utility):<br>    j ← GetRandomNode()<br>    j.GetState(i)<br>    if i.Utility ≤ j.Utility<br>        CopyStatePartial(j)<br>        Mutate(i)<br>    Utility ← 0.0 (reset utility) | j ← this node<br>do forever:<br>sleep until a request received<br>GetState(i) – send j state to node i:<br>    Send j.Utility to i<br>    Send j.Links to i<br>    Send j.Strategy to i |
| Function CopyStatePartial(j): | Function Mutate(i): |
| i.Strategy ← j.Strategy<br>drop each link from i with prob. W<br>for each link in j.Links:<br>    i.addLink(link) | with prob. M mutate i.Strategy<br>with prob. MR mutate i.Links:<br>    drop each link with prob. W<br>    i.addLink(SelectRandomNode()) |

Figure 1. The SLACER protocol runs continuously in each node. It is composed of an active and passive thread. The active thread is constantly executed; the passive thread wakes only to serve requests for information from other nodes. The GetRandomNode() function supplies a random node from the entire population. Three probabilities: W, M, MR defines a space of protocols. When W = 1 the protocol collapses to the previous SLAC protocol, of which SLACER is an extension

Periodically, each node ($i$) compares its performance against another node ($j$), randomly selected from the population. If $Ui \leq Uj$ then node $i$ drops each of its current links to other nodes with high probability $W$, and copies all node $j$ links and adds a link to $j$ itself. Additionally $i$ then copies $j$'s strategy – the strategy represents some behaviour that nodes execute during application level interaction. After such a copy operation has occurred, then, with low probability $M$, node $i$ adapts its strategy and with probability $MR$ adapts its links. Adaptation involves the application of a "mutation" operation. Mutation of the links involves removing each existing link with probability $W$ and adding a single link to a node randomly drawn from the network. Mutation of the strategy involves applying some form of change in application behavior with probability $M$ - the specifics of strategy mutation are dictated by the application domain (see later). After the periodic utility comparison, whether this resulted in the copying of another node or not, the node resets its utility to zero - hence, utilities do not accumulate indefinitely.

Each node is limited to a maximum number of links or neighbors (its so-called view size). If any SLACER operation causes a node to require an additional neighbor above this limit then a randomly selected existing link is removed to make space for the new link. Links are always undirected and symmetrical, so that if node *i* links to node *j*, then *j* must also maintain a link to node *i* and conversely if node *i* breaks a link to node *j* then node *j* also breaks its link to node *i*. For implementation purposes, SLACER requires additional passive thread functions (not shown in figure 1) that would handle "addLink and dropLink" requests from other nodes.

Previous "tag" models, on which SLACER is based (Hales 2000) have indicated that the rate of mutation applied to the links needs to be significantly higher than that applied to the strategy by about one order of magnitude hence $MR >> M$.

When applied in a suitably large population, over time, the algorithm follows an evolutionary process in which nodes with high utility replace nodes with low utility. However, as will be seen, this does not lead to the dominance of selfish behavior, as might be intuitively expected, because a form of social incentive mechanism results from the emergent network topology (a friendship network). This means that high utility but anti-social strategies, even though favored by the individual nodes, do not dominate the population. The topology therefore guides the adaptation of the strategy away from anti-social behaviors.

### 3.1 Random Sampling via NEWSCAST

In Figure 1, we assume a function that returns a random node from the entire population of nodes irrespective of the current network topology. This function cannot use the network maintained by SLACER itself since it can become partitioned. In our simulations we used the existing NEWSCAST algorithm (Jelasity et al 2004) in order to provide this service. NEWSCAST provides exactly this function by maintaining *its own* scalable and robust random overlay network based on a gossip protocol – in which random neighbors constantly gossip their network views (the links they maintain to other nodes). NEWSCAST maintains a random and fully connected topology even under conditions of high node failure and malicious behavior. Since NEWSCAST maintains its own overlay it can be deployed to support SLACER in a modular way – the only interaction necessary is via a GetRandomNode() function invoked from SLACER to the lower newscast layer and returning a randomly sampled node of the network.

### 4. The Prisoner's Dilemma Test Application

The two player single-round Prisoner's Dilemma (PD) game captures, in an abstract form, a situation in which there is a contradiction between collective and individual self-interest. Two players interact by selecting one of two choices: to "cooperate" (C) or "defect" (D). For the four possible outcomes of the game, players receive specified payoffs. Both players receive a reward payoff (R) and a punishment payoff (P) for mutual cooperation and defection respectively. However, when individuals select different moves, different payoffs of temptation (T) and sucker (S) are awarded to the defector and the cooperator respectively. Assuming that neither player can know in advance which move the other will make and wishes to maximize her own payoff, the dilemma is evident in the ranking of payoffs: $T > R > P > S$ and the constraint that $2R > T + S$. Although both players would prefer T, because its the highest payoff, only one can attain it in a single game. No player wants S because it's the lowest payoff. No matter what the other player does, by selecting a D-move a player ensures she gets either a better (T) or an equal (P) payoff to her partner. Playing D, then, can't be bettered since this ensures that the defector cannot be suckered (S). As any modern Economist would no doubt tell you, this is the so-called "Nash" equilibrium for the single round game – hence an ideally rational selfish player would always choose D. It is also an evolutionary stable strategy (ESS) for a population of randomly paired players, where reproduction fitness is proportional to payoff, so Darwin can't save us from the Economists this time!

Therefore, the dilemma is that if both individuals selected a cooperative (C) move they would be jointly better off (getting R each) than if they both select D, but both evolutionary pressure and ideal rationality result in mutual defection, so the players only get P each, because there is always an individual incentive to select defection.

We select this game as a minimal abstract basis for a test "application" that captures, abstractly, a whole range of possible application tasks in which nodes need to establish cooperation and trust with their neighbors but without central authority or appeal to external mechanisms. The PD payoff structure gives each node an individual incentive to defect while specifying a social (joint) incentive to cooperate. We can relate this to a P2P application task where individual nodes need to behave in a non-selfish way to increase network level performance. This could include altruistically sharing files and resources or passing on a message, to facilitate communication between a sender and receiver node, or using resources warning friends about a virus program and supplying them with a fix when you've already inoculated yourself against the infection.

Our test application involves all nodes playing the PD with randomly selected neighbors in the SLACER constructed social network. A node can only choose one of two strategies – cooperate or defect. The utility value required by SLACER is then set by the PD application as the average payoff the node received from these game interactions. The SLACER algorithm then adapts the links and strategy of the nodes as discussed previously.

In past work, using the SLAC algorithm, we found that results form a similar PD test application were good indicators of results obtained when the same algorithm was tested in a more realistic P2P file-sharing scenario (Hales et al 2005) – where free-riding (or leeching) was the analogue of defection and altruistically sharing files represented cooperation. We therefore use the PD here as a minimal or canonical test from which initial evaluations of SLACER can be obtained.

**5. Measuring Cooperative Connected Paths (CCPs)**

Our goal in designing SLACER is to self-organize artificial social networks (ASN) with some of the disable properties we observe in human social networks that would be of value in many P2P application domains. These desirable properties comprise: short paths between all members, high cooperation between neighbors and paths of cooperation between non-directly connected members.

In order to evaluate the quality of the produced ASNs we measured, over the entire population of nodes: average path length, proportion of cooperator nodes and the proportion of nodes linked via chains of cooperators. For the latter we introduce a Cooperatively Connected Path (CCP) measure that quantifies the amount of "cooperative connectedness" in an ANS.

Given a population of nodes in which each node is in one of two states (cooperate or defect) we define the CCP measure as the proportion of all possible node pairs that are either linked directly or for which there exists at least one path between them in which all intermediate nodes are in a cooperative state. This definition does not require that the node pairs themselves are cooperative, just that a cooperative path exists between them, or a direct link. See figure 2 for an example.

We can illustrate this concept with a hypothetical message passing game in a human social network. Suppose I wish to pass a confidential message to a distant individual to which I am not directly known (the President of the United States, say) and I want the message to arrive without being compromised (read or changed). It would not be possible to achieve this if I could not find a route of high trust and cooperation through my social network from myself to the final recipient. That is, I would need to pass the message via a set of intermediaries who could be trusted to act cooperatively. However, this task does not require that either the sender or final recipient behave cooperatively if they are passed messages for others.

In a connected network in which all nodes were cooperators, the CCP measure would obviously be 1. However, even with a number of defector nodes inhabiting such a network the CCP can still be 1, so long as the defectors were located in such a way that they do not obstruct other cooperative routes between pairs of nodes – i.e. there are alternative cooperative routes around the defectors. Hence, the CCP measure is determined by the topology of the network *and* the strategy distribution over nodes in that topology (the strategic topography).

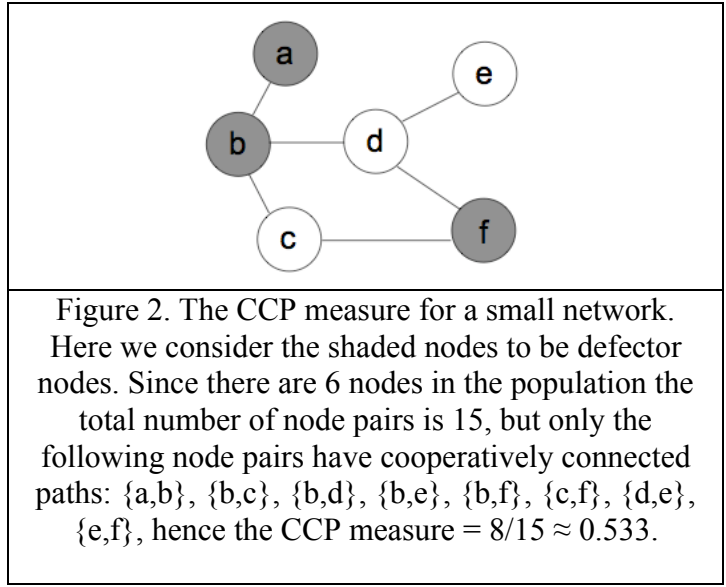

Figure 2. The CCP measure for a small network. Here we consider the shaded nodes to be defector nodes. Since there are 6 nodes in the population the total number of node pairs is 15, but only the following node pairs have cooperatively connected paths: {a,b}, {b,c}, {b,d}, {b,e}, {b,f}, {c,f}, {d,e}, {e,f}, hence the CCP measure = 8/15 ≈ 0.533.

By combining the CCP measure with other measures, we can get an idea of the strategic topography of the ASN. For example, if the network was connected (not partitioned) and the proportion of cooperating nodes was high but the CCP was low, then we would conclude that a few non-cooperative nodes occupy positions in the network that allow them to block a large number of unique cooperative paths between nodes.

Going back to our design goal for SLACER, our intention is to generate connected networks of cooperative nodes, so our ideal outcome would be ASN with CCP at or close to the maxim of 1.

**6. Simulation Specifics**

We performed simulation experiments using two independent implementations and verified the results matched (we "docked" the models). We did this to minimize the effect of possible implementation artifacts. All the observations and conclusions we draw are supported by the results obtained from both implementations[6].

**6.1 One Time Cycle**

When nodes engage in their application task this involves playing the PD with randomly selected neighbors then nodes periodically initiate the "compare utility" activity in the active thread of the SLACER algorithm. We programmed the simulations so that, on average, over one cycle, each node initiates some application level activity, causing a utility update and executes one "compare utility" call. We simulated this both semi-asynchronously and fully asynchronously.

---

[6] An implementation in the PEERSIM environment is available at: http://peersim.sourceforge.net

In one time cycle, 10N nodes are selected from the population randomly with replacement. When a node is selected, it chooses a neighbor at random and plays a single-round game of the PD. If a node is selected that currently has no neighbors then a link is made to a randomly selected node so a game can be played. A game is played by both nodes executing the strategy indicated by their current state (cooperate or defect). Both nodes then increment their utility by the appropriate payoff. If both cooperated then both increase their utilities by R, if both defect then both increment by P, otherwise the cooperating node increments by S and the defecting node gets T. When nodes compare utility in the SLACER algorithm the average utility over the total number of games played is used rather than the absolute utility.

**6.2 Two Simulation Variants**

In the semi-asynchronous simulation implementation, after nodes had played their PD games, N nodes were selected, randomly, with replacement, to execute the "compare utility" operation (see Figure 1). A new cycle was then started. In the fully asynchronous version the same operation was executed probabilistically by each node after each PD game was played – with probability 0.1.

What we model here is the notion that for some possible application tasks utility updates may be instantaneous and asynchronous, for example, neighbor nodes may be able to provide requested resource immediately in some application tasks. However, in others rewards may be delayed and synchronized over a large set of nodes – especially when a number of nodes need to coordinate together to achieve a collective task and only then get their reward – consider a P2P search engine pooling its page rankings, for example. Obviously many different schemes are possible which would be dictated by the specifics of an application task but for the purposes of the initial PD test application, we considered these two.

**6.3 Parameter Settings**

We set the PD payoffs to T = 1.9, R = 1, P = 2d, S = d, where d is some small value (in this case $10^{-4}$). We set the mutation rates to M = 0.001 and MR = 0.01. Nodes were allowed a maximum view size of 20 (hence each node could link to maximum of 20 neighbors). We ran simulation with the rewire probability W = 0.9 and W = 1 over a range of network sizes from N = 2,000 to 64,000.

**7. Simulation Results**

In our simulation experiments, we initialized the entire population of N nodes to the state of defection and connected them in a random network topology (we found, in fact, that any initial network topology, including fully disconnected, had no significant effect on the outcomes of the simulations). We then measured how long it took in time cycles for high levels of cooperation to emerge - we stopped when 98% of all nodes were in a cooperative state. We found cooperation and topology measures meta-stabilized after this point, randomly oscillating around mean values. Complete stability is not possible because nodes always continue to move and change strategies within the network.

When a state of high-cooperation was reached, we measured the CCP for the entire population along with the average path length, the clustering coefficient and the largest single connected component. For each experiment, we performed a number of runs (10) with identical parameters but different pseudo-random number generator seeds. In our analysis we took into account the averages and variances over these runs.

We compared two "rewire" probabilities (see figure 1) W = 0.9 and W = 1. When W = 1, this eliminates completely the storage of old links when a node copy occurs and this collapses the algorithm to the previous SLAC algorithm – which we know, from previous work, produces highly partitioned, though cooperative, networks.

**7.1 Cycles to High Cooperation**

Figure 3 shows the number of cycles to high cooperation from an initial population of all defector nodes. As can be seen SLACER takes longer than the original SLAC algorithm to converge to high cooperation but only by a few tens of cycles. It is an important test to ensure a network can recover high cooperation from total defection quickly, since this ensures robustness against a possible catastrophic failure in cooperation caused by other means. We have an existence proof that the algorithm *can* escape from this dire situation quickly. We got similar results to Figure 3 when we initialized the nodes without any links at all; this indicates the ability of SLACER to recover from a total network outage quickly.

**7.2 SLAC Networks**

Figure 4 shows the size of the largest connected component and the CCP value when W=1, for various network sizes, after high cooperation emerges. The largest component size does not grow as the network size increases and hence this degrades the CCP value. The CCP value starts small and does not scale. This is because the nodes are partitioned into numerous small components; although nodes are cooperative to their neighbors, the partitions stop cooperative paths being made between nodes in different components. The SLAC algorithm, then, leads to ASN that suffer from something that might be termed "extreme tribalism". It's rather like the human equivalent of a population composed entirely of self-absorbed cliques or cults that, although internally cooperative, want nothing to do with each other.

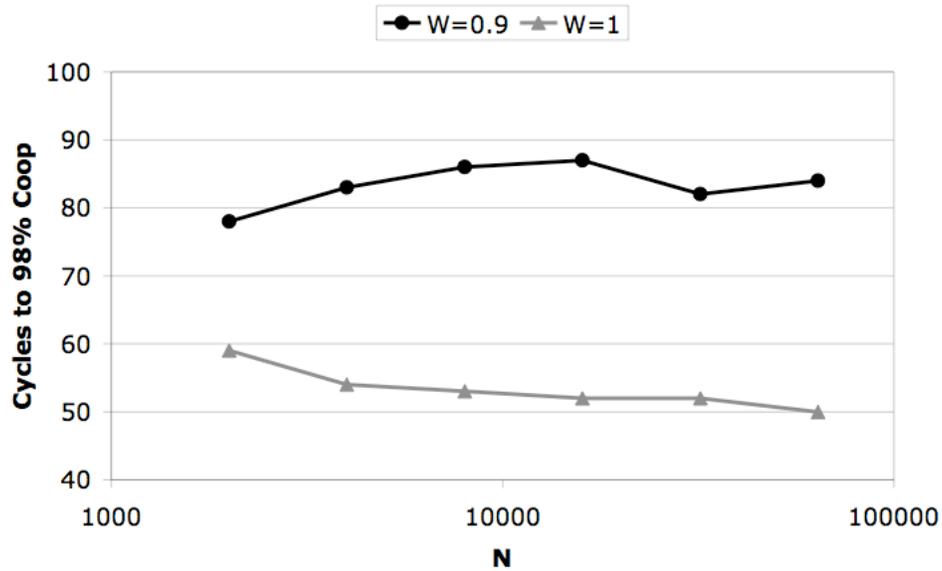

Figure 3. Average number of cycles to reach high cooperation for various network sizes (N). Values are shown for rewire probability W=0.9 (SLACER) and W=1 (SLAC). Each point is an average of 10 runs. The variation over individual runs was never more than about 10 cycles either side of the mean. As can be seen, on average, it takes no more than an additional 40 cycles for high cooperation to be achieved when W=0.9.

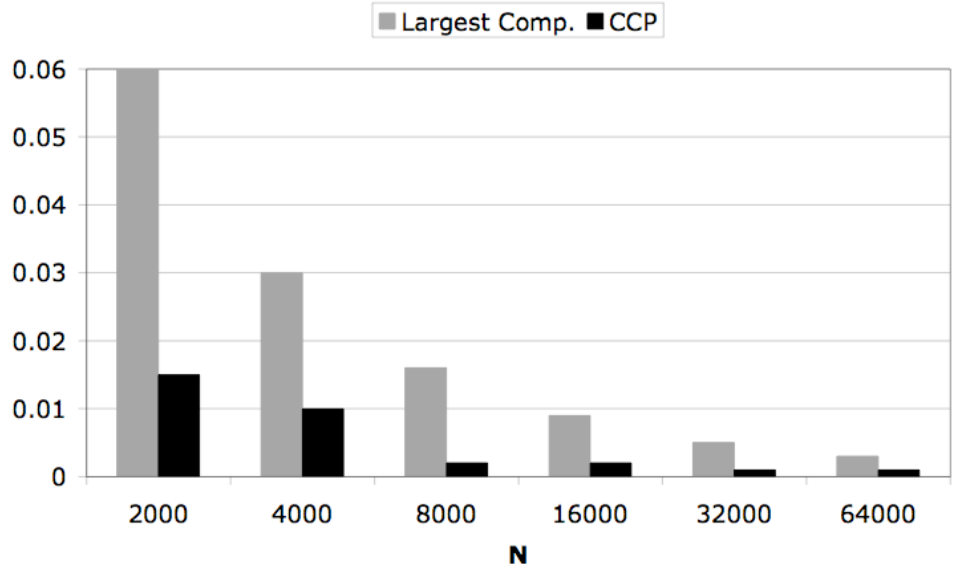

Figure 3. The size of the largest component as a proportion of population size and the Cooperative Connected Path (CCP) measure for W=1 (the SLAC algorithm) after high cooperation has been achieved. Each bar shows an average of 10 runs - variance was negligible. The curve for the largest component indicates that its absolute size remains approximately constant (about 150 nodes).

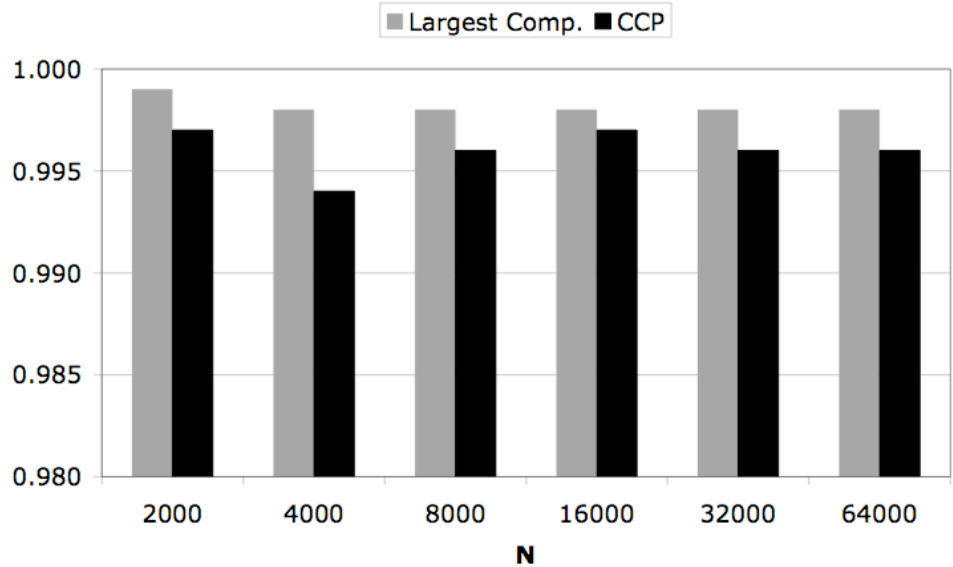

Figure 5. The size of the largest component as a proportion of population size and the Cooperative Connected Path (CCP) measure when W=0.9 (SLACER). Each bar shows an average of 10 runs. It can be seen that almost all nodes inhabit a giant connected component (GCC). The high CCP indicates that the vast majority of nodes within the GCC are connected by cooperative paths. These results can be compared to Figure 4.

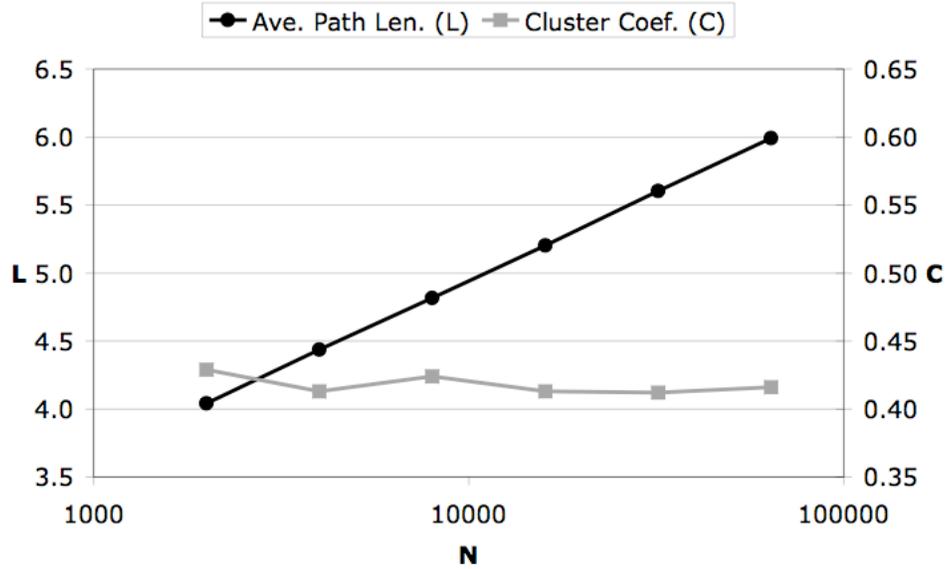

Figure 6. The clustering coefficient (C) and average path length (L) when W=0.9 and after high cooperation is attained for different network sizes (N). Note that while C stays more-or-less constant, L increases log-linearly with N. Each point is an average value of 10 simulation runs. The variance of results was negligible.

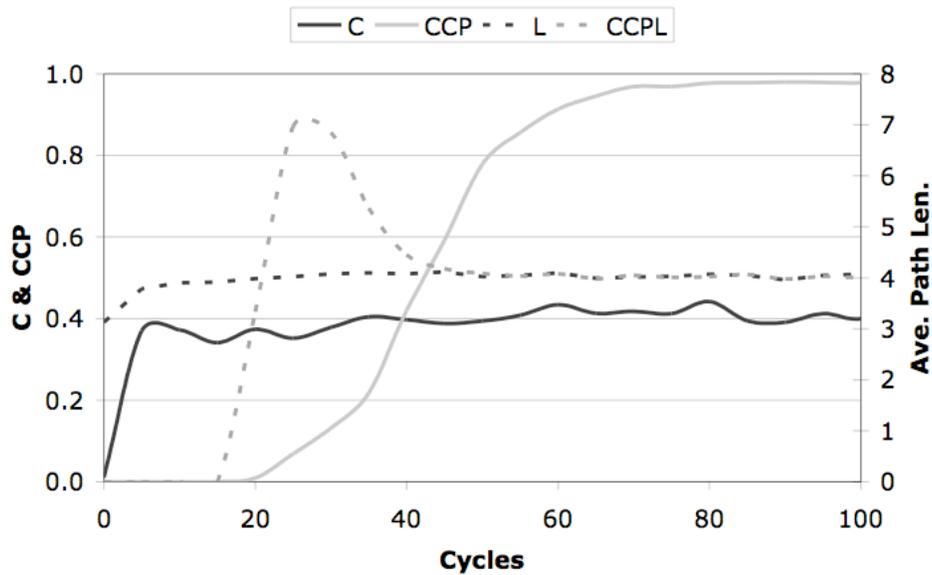

Figure 7. A typical individual simulation run for a 2000 node network. Shown are the Clustering Coefficient (C), the average path length (L), The Cooperative Connected Paths (CCP) measure and the average Cooperative Connected Path Length (CCPL). This latter measure gives the average path length of cooperative connected paths. Notice that it peaks as CCP starts to rise then settles down to agreement with L. This shows that when cooperation first emerges it spreads quickly over the network – forming a kind of extended "cooperative backbone" which soon seeds the entire population, reducing the CCPL to the L value.

### 7.3 SLACER Networks

The results in figure 4 can be compared to those given in figure 5, in which W=0.9. Here we see very high values for both CCP and the largest component size. Almost all nodes inhabit a giant connected component (GCC) that, although containing some defecting nodes, provides cooperative routes between the large majority of its members. SLACER therefore generates cooperative ASN without the "extreme tribalism" produced by SLAC. Importantly, the CCP scales well because as N increases CCP does not decrease. This gives us confidence that SLACER would produce high quality ASN for populations of any practical size.

### 7.4 Basic Topological Features

In figure 6, we measure some other structural characteristics of the ASN produced by SLACER. Again, we take these measures over various network sizes after high cooperation has been achieved so the results are directly comparable to the previous figures. When we plot the average path length (L) and clustering coefficient (C) we see a number of interesting results. Firstly, the ASN follow a "small-world like" topology, since C is relatively high (compared to a random network) yet L is low – meaning most nodes are connected by only a few hops. We note that L scales-up log-linearly, indicating that even in very large networks most nodes are connected by short paths. The degree distributions (not shown) of the networks appear linear with about 10% of nodes having the maximum number of links (in our case 20, the view size) and almost no nodes holding zero links. This means that many nodes have many links. This suggests the ASN are more robust than some human social networks which display power-law distributions, giving a scale-free topology, where only a few hub nodes have many links.

### 7.5 Typical Evolution

Figure 7 shows a time series of a typical run for a 2,000 node network executing SLACER (W=0.9). We can identify a number of distinct stages in the time evolution of the network before high cooperation is reached. First, the clustering coefficient (C) increases rapidly and the path length (L) increases prior to cooperation forming. This results from a randomized rewiring process, since all utilities are identical. This is sufficient to create the C and L values we find throughout the run. Just before cycle 20, via random mutation, two linked nodes become cooperative. This "seed tribe", rather than growing locally, explodes - creating a kind of sparse "cooperative backbone" over the entire network. This spreads cooperation quickly, leading to rapid saturation of cooperation. These four stages; random rewire, seed formation event, seed explosion then saturation – we observe in all runs we have examined.

We have discussed in detail, in previous works (Hales 2000, 2004, Hales et al 2005), the basic nature of this process and how, it operates: though a process of "group like" selection between clusters or "tribes". This could be termed "tribal selection". Essentially, if tribes are organized such that they provide high utility to their members then they will tend to recruit more members. Tribes that become "infiltrated" with defectors will tend to die out – since nodes will move to the tribes offering better utility. Ironically, by defecting and acting selfishly a node sows the seeds of its own tribe's destruction since it's initial high utility leads to it becoming surrounded by copy-cat defectors reducing it's payoffs.

**7.6 Robustness to Churn**

We have also subjected the ASN produced by SLACER to robustness tests by introducing various amounts of "churn" - where old nodes leave and new nodes join the network. In these experiments, we reset randomly selected nodes to defect states over various intervals of cycles. We found that even when 50% of nodes were replaced at one cycle then high-cooperation, CCP and the topology structures previously observed quickly reformed within a few cycles. This was expected, since SLACER incorporates noise in the form of mutation to both links and strategies, driving its evolutionary dynamics and, as we have shown, can quickly recover from states of complete defection and link disconnection.

**7. 7 Different Rewire Values**

We have also experimented with different values of W. We found that with values of W higher than 0.9 the CCP started to fall off and we were back to the extreme tribalism of SLAC. However, when we reduced W below 0.9 we found that the amount of cooperation in the nodes began to fall, as did the clustering coefficient (C), so that when W = 0.7 cooperation levels above 90% of nodes were never achieved (with mean values oscillating around 80%) and C at around 2.5. For W = 0.5, cooperation levels never got above 70%, oscillating widely around 60% with C below 2. This indicates a trade-off; we need high W to get high cooperation – because this creates the clusters or "tribes" that drive the process – however, if W is too high we get "extreme tribalism" – i.e. a disconnected network. Varying W, therefore, controls the strength of tribalism or "cliqueyness".

**8. Conclusion**

We believe that SLACER is a step towards potentially *very* useful ASN. We are confident that we can apply it to more realistic and useful application tasks since we have had earlier success applying the SLAC algorithm to a P2P file-sharing application and we should be able to follow the same approach.

We have a specific interest in producing ASN for collective P2P virus, spyware and spam filtering. Another possible target application is supporting trust and cooperative interactions between nodes in a P2P search engine. In all these cases, although sophisticated P2P algorithms *already exist*, they tend to assume all peers will act in a cooperative way, which is naive and unsafe in open systems, or they require *pre-existing trusted social networks* to be supplied as input – which are not always available or appropriate.

SLACER can plug this gap with a self-organizing, robust, dynamic and modular algorithm producing ASN, tuned, in real time, directly for specific applications. This requires that already existing P2P algorithms be adapted such that they periodically calculate a utility value, indicating the current service and trust quality derived from the network and provide some adaptation method allowing for different levels of service and trust to be offered to the network.

We have assumed that nodes will implement SLACER correctly, sending correct utility values, links and strategies. It could be argued that by assuming this we are merely *begging the question*, falling into the naive trap of assuming the trust and cooperation that we claim to create! However, we always have in mind, that the determination of the properties of other nodes could follow an indirect method to reduce malicious noise – by, for example, asking some set of neighbors of a node to verify the information sent or using other schemes. Additionally, utilities can be reduced to a binary satisfaction function that obviates the need for utility comparisons or strategy copying completely. Then nodes move in the network and change their strategies if they are not satisfied (Singh et al 2005). The fact that SLAC works when nodes *are* able to copy those with higher utility indicates that it is robust when nodes are able to scan the population and select those behaviors that are producing high individual returns – irrespective of their collective effect. In any implemented open system, such behavior is possible if nodes collusively exchange information.

Where applications are complex, involving many kinds of trust and interaction tasks between peers, we envisage the creation of multiple instantiations of SLACER each supporting different tasks running concurrently. In this way, each ASN would be tuned to the particular requirements of the specific task associated with it - rather like the way humans form many networks around the different goals and tasks they need to achieve in everyday life.

Could it be that by teaching our computers to make friends with each other we could finally wave goodbye to the much of the anti-social behavior that plague Internet? On the other hand, can even humans know who their true friends really are?

## Acknowledgements

We thank Ozalp Babaoglu, Mark Jelasity, Alberto Montresor and Simon Patarin from the University of Bologna, Dept of Computer Science, for perceptive discussions, observations and pointers concerning P2P systems. We also thank Andrea Marcozzi and Gian Paolo Jesi, who produced an initial implementation of SLAC using NEWSCAST in the PEERSIM environment.